# Estimating the resolution of real images

**Ryuta Mizutani[1]\*, Rino Saiga[1], Susumu Takekoshi[2], Chie Inomoto[2], Naoya Nakamura[2], Makoto Arai[3], Kenichi Oshima[3], Masanari Itokawa[3], Akihisa Takeuchi[4], Kentaro Uesugi[4], Yasuko Terada[4] and Yoshio Suzuki[5]**

[1] Department of Applied Biochemistry, Tokai University, Hiratsuka, Kanagawa 259-1292, Japan
[2] Tokai University School of Medicine, Isehara, Kanagawa 259-1193, Japan
[3] Tokyo Metropolitan Institute of Medical Science, Setagaya, Tokyo 156-8506, Japan
[4] Japan Synchrotron Radiation Research Institute (JASRI/SPring-8), Sayo, Hyogo 679-5198, Japan
[5] Graduate School of Frontier Sciences, University of Tokyo, Kashiwa, Chiba 277-8561, Japan

\*Correspodence: ryuta@tokai-u.jp

**Abstract**. Image resolvability is the primary concern in imaging. This paper reports an estimation of the full width at half maximum of the point spread function from a Fourier domain plot of real sample images by neither using test objects, nor defining a threshold criterion. We suggest that this method can be applied to any type of image, independently of the imaging modality.

## 1. Introduction

The resolutions of X-ray microtomographic images have been estimated by using square-wave patterns [1]. Such patterns should be prepared with a precision much finer than the target resolution. Another method of resolution estimation is to observe edge profiles [2]. However, edge profiles can be distorted with the refraction effect, resulting in overestimation of image resolvability. A modulation transfer function (MTF) representing the frequency domain response can be determined from square-wave patterns or edge profiles. The MTF has been used to estimate the resolution, though a threshold must be artificially defined in order to specify the MTF cutoff.

The resolution can also be estimated from the high frequency limit of the Fourier domain profile. It has been proposed that five times the noise level in amplitude (Rose criterion) or twice the noise level in the power spectrum [3] should be used to define the high frequency limit. The resultant estimates largely depend on the threshold criterion [4] and hence involve arbitrariness.

This paper reports an estimation of the full width at half maximum (FWHM) of the point spread function (PSF) from a Fourier domain plot of real sample images by neither using test objects, nor defining a threshold criterion. The resolutions of micro/nano-tomographic sections were also estimated. We suggest that this method can be applied to any image, independently of the imaging modality.

## 2. Resolution estimation from a Fourier domain plot

The FWHM of the PSF can be estimated from a Fourier domain plot, as reported by Mizutani *et al.* [4]. This method is based on the principle that a Gaussian at an arbitrary position in the real domain is transformed into a Gaussian at the origin in the Fourier domain. The FWHM of the transformed Gaussian is then determined by plotting the logarithm of the squared norm of the Fourier transform against the squared distance from the origin. This method is only effective when the PSF can be approximated with a Gaussian, though the Gaussian approximation is applicable in most cases. Figure 1 shows example plots calculated from test images.

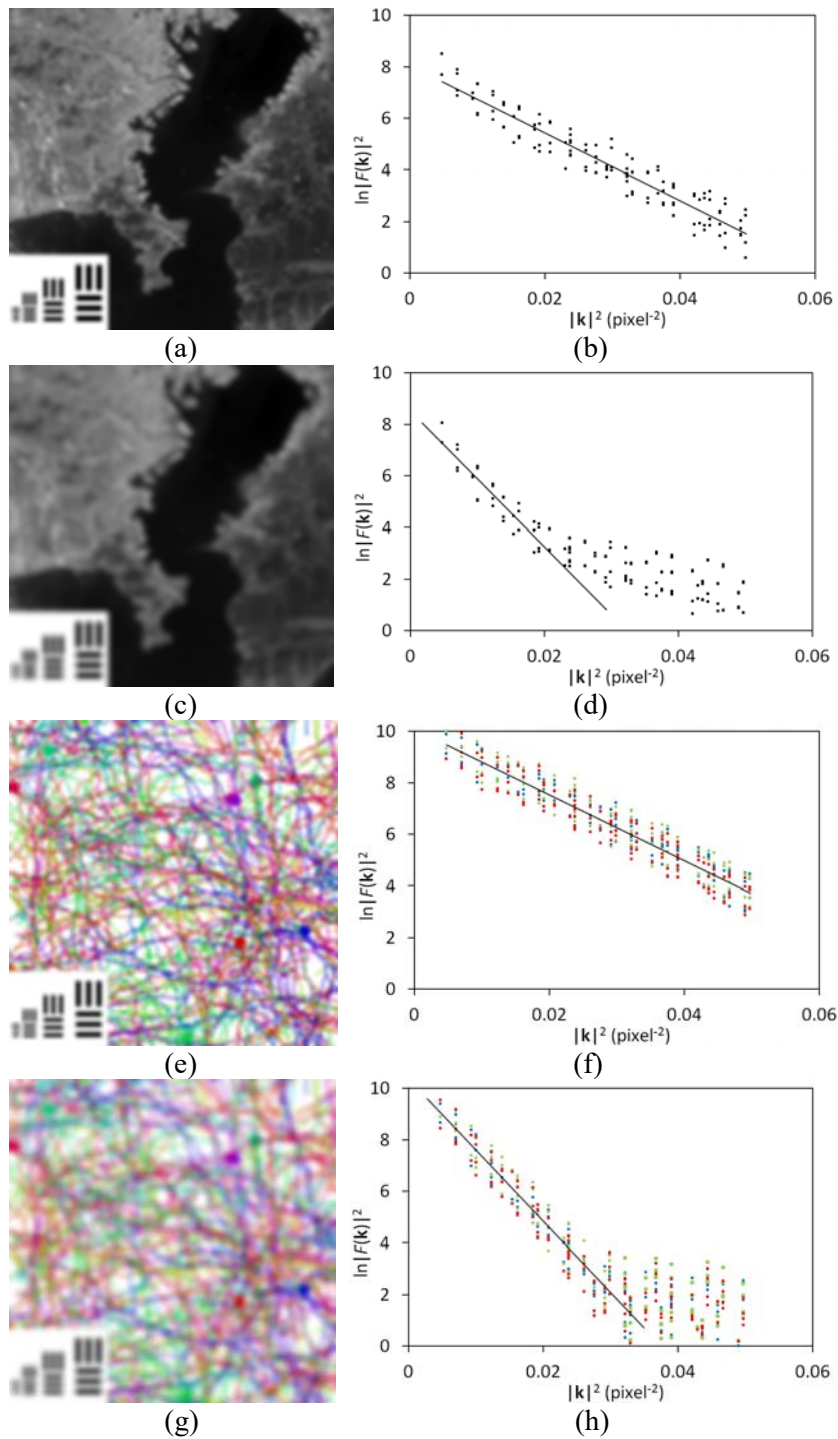

**Figure 1.** A satellite image of Tokyo Bay (a, c) and a color drawing of a human brain network (e, g) were convolved with Gaussians having FWHMs of 4 pixels (a, e) and 6 pixels (c, g). The lower left inset showing square wave patterns with 2, 4, 6 and 8-pixel pitches indicates that blurring comparable to the Gaussian FWHM occurred. Fourier plots of Mizutani *et al.* [4] were calculated from these test images (b, d, f, h). Red, green or blue dots indicate plots of each color plane. Linear correlations representing Gaussian PSFs are indicated with thin lines.

A satellite image of Tokyo Bay (JAXA) was converted into gray-scale and subjected to 2 × 2 binning in order to eliminate intrinsic blurring. A color drawing was taken from a figure of the human brain network [5]. Square wave patterns with pitches of 2, 4, 6 and 8 pixels were embedded in these original images. Then, test images were generated by convolving the original images with Gaussians. Red, green and blue planes of the network drawing were individually processed.

The resolution estimation procedure was implemented in the RecView software (available from https://mizutanilab.github.io/). Regions within 10 pixels from each Fourier coordinate axis were not used for the plot, since streaks were observed along the axes in some images, presumably due to the readout properties of the detector. Then, pixels in each 5 × 5 bin were averaged in order to reduce data and simplify the plot. In this plot, the Gaussian PSF should be observed as a linear correlation [4]:

$$\ln|F(\mathbf{k})|^2 \cong -4\pi^2\sigma^2|\mathbf{k}|^2 + \text{constant},$$

where $F(\mathbf{k})$ is the Fourier transform of the image, $\mathbf{k}$ the coordinate in the Fourier domain, and $\sigma$ the standard deviation of the Gaussian PSF. The equality is true when the images can be regarded as random, though most images seem to meet this requirement. Indeed, plots generated from test images showed linear correlations corresponding to the FWHM of the Gaussian PSF (Figure 1), independently of the image types. These results indicate that this method can be used to estimate the image resolution.

## 3. Resolution estimation of X-ray micro/nano-tomographic sections

The resolution of an X-ray micro/nano-tomographic section depends not only on the optics performance but also on a number of factors, including rotation stage wobbling, component drift, vibration, temperature fluctuation, and sample deformation during the data acquisition. Therefore, the resolution must be determined using tomographic sections and not from a single raw image. Figure 2a shows a microtomographic section of test patterns carved on an aluminum wire [2]. This image was reconstructed from a dataset acquired at the BL20XU beamline of SPring-8 with a simple projection geometry using a LuAG scintillator screen. The pixel width was 0.50 μm. The pattern with a pitch of 1.60 μm is clearly resolved, while the pattern with a 1.20 μm pitch is not visible.

Figure 2b shows the Fourier domain plot generated from the section shown in Figure 2a. In this case, the PSF slope was estimated from the linear regression between 0 and 0.05 pixel$^{-2}$, giving a slope of -48.4 pixel$^{-2}$. The FWHM of the PSF was calculated as $2\sqrt{2\ln 2} \times \sqrt{-\text{slope}}/2\pi$, giving a FWHM of 2.61 pixels. By multiplying the FWHM with the pixel width, the resolution was estimated to be 1.30 μm. This coincides with the appearance of the test patterns (Figure 2a).

Figure 2c shows a nanotomographic section of a human neuron [4]. The brain tissues were collected with informed consent from the legal next of kin using protocols approved by ethical committees of the related organizations. The image was reconstructed from a dataset acquired at the BL37XU beamline of SPring-8 using FZP optics. The effective pixel width was 23 nm. The plot of this section (Figure 2d) showed a typical profile that can be divided into three parts: the origin spike, successive PSF slope, and the right remainder of shallow descent.

The origin spike is ascribable to the Fourier profile of the overall structure. The region near the origin in the Fourier domain corresponds to large structures in the real domain. If a sample occupies only small part of the viewing field, the spike becomes more prevailing and distorts the PSF slope. In such cases, the region of interest instead of the whole image should be used to estimate the resolution.

Although the right shallow slope is distinct in Figure 2d, this slope should not be used to estimate the resolution, as discussed below. In this case, the PSF slope should be taken from the left quarter of the plot. The resolution estimated from the linear regression between 0 and 0.01 pixel$^{-2}$ was 149 nm, while the resolution of the same nanotomograph evaluated using a test object was 100-120 nm [4]. The difference would originate from tissue deformation or sample drift during the data acquisition.

The right shallow descent represents the FWHM of 2 pixels and corresponds to the Nyquist limit of digitization. Therefore, the shallow descent must not be used to estimate the resolution. The signal-to-noise ratio of the neuron section (Figure 2c) is much lower than that of the test object section (Figure 2a). This is mainly because the pixel width of Figure 2c is approximately 22 times smaller than that in

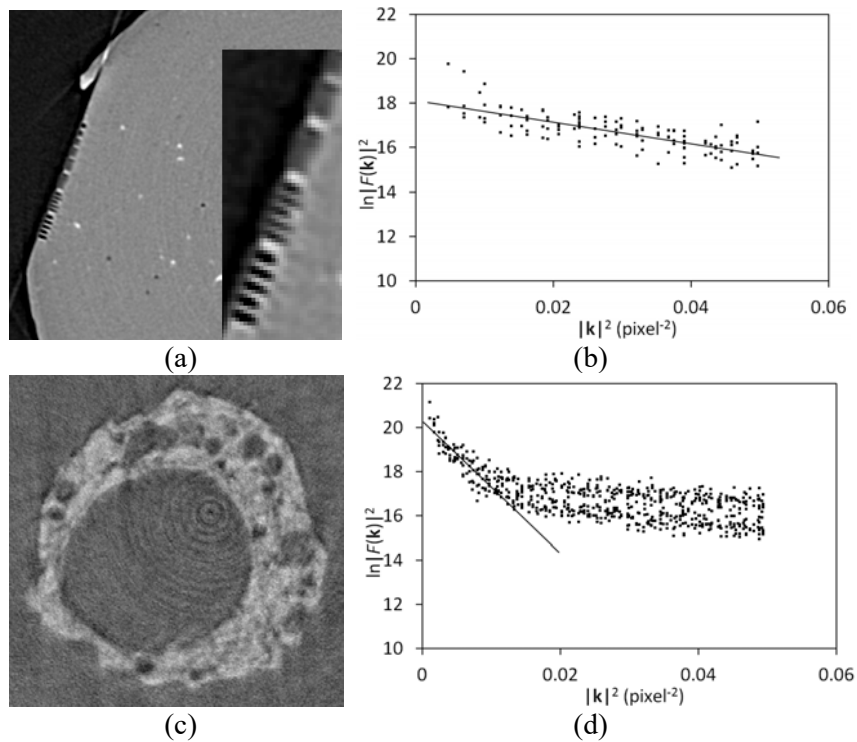

**Figure 2.** (a) Microtomographic section of an aluminum test object. The right inset shows magnified patterns. Patterns with pitches of 2.0 and 1.6 μm were clearly resolved, while those of 1.2 μm and lower were not. (b) Fourier domain plot of the test object section. The thin line indicates the possible PSF slope. (c) A nanotomographic section of human brain neuron. Image width: 10 μm. (d) Fourier domain plot of the neuron section. The thin line indicates the possible PSF slope. The right remainder slope corresponds to the Nyquist limit of digitization.

Figure 2a. Noise in the image raises the Nyquist slope, making it difficult to identify the PSF slope, as can be seen in Figure 2d. Hence, raw data should be acquired with a sufficient signal-to-noise ratio to visualize fine structures, and in turn, to allow the detection of the PSF slope.

The results obtained in this study indicated that the FWHM of PSF can be estimated from the sample image itself. We suggest that this method can be applied to any type of image, independently of the imaging modality.

**Acknowledgments**
This work was supported in part by Grants-in-Aid for Scientific Research from JSPS (nos. 21611009, 25282250 and 25610126). The synchrotron radiation experiments were performed with the approval of JASRI (proposal nos. 2013A1384, 2014B1083, 2015A1160, 2015B1101 and 2016B1041).

**References**
[1] Mizutani R, Takeuchi A, Uesugi K and Suzuki Y 2008 *J. Synchrotron Radiat.* **15** 648
[2] Mizutani R, Taguchi K, Takeuchi A, Uesugi K and Suzuki Y 2010 *Nucl. Instrum. Meth. A* **621** 615
[3] Modregger P, Lübbert D, Schäfer P and Köhler R 2007 *Physica Status Solidi (a)* **204** 2746
[4] Mizutani R *et al* 2016 *J. Microsc.* **261** 57
[5] Mizutani R, Takeuchi A, Uesugi K, Takekoshi S, Osamura R Y and Suzuki Y 2010 *Cerebral Cortex* **20** 1739